\documentstyle[12pt]{article}
\def\be{\begin{equation}}
\def\ee{\end{equation}}
\def\bea{\begin{eqnarray}}
\def\eea{\end{eqnarray}}
\def\a{\alpha}
\def\b{\beta}
\def\g{\gamma}
\def\d{\delta}
\def\G{\Gamma}
\def\m{\mu}
\def\n{\nu}

\author{M.V. Gorbatenko,   A.V. Pushkin,  H.-J. Schmidt}

\title{On a relation between the Bach equation and the equation of 
geometrodynamics}

\date{}
\begin{document}
\maketitle

\begin{abstract}
The  Bach equation and the  equation  of geometrodynamics are based on 
two quite 
different physical motivations, but  in both  approaches, the  conformal
properties of 
gravitation plays  the key role. In this paper we present an analysis of 
the relation 
between these two equations and show that the solutions of  the equation 
of geometrodynamics  are of a  more general nature. We show the following  
non-trivial 
result: there exists a conformally invariant Lagrangian, whose field  
equation
generalizes the  Bach equation and has  as solutions those Ricci 
tensors which  are solutions to the equation of geometrodynamics.
\end{abstract}

\bigskip

\bigskip

\noindent
Addresses of the authors: 

\bigskip
\noindent
{\small {Russian Federal Nuclear Center -
All-Russian S\&R Institute of Experimental Physics. Sarov, Nizhnii 
Novgorod  Region, 607190, Russia. \par 
\noindent 
E-mail: gorbatenko@vniief.ru \quad and \quad  pav@vniief.ru \quad resp.

\bigskip

\noindent 
Author for correspondence: H.-J. Schmidt, Institut f\"ur 
Mathematik, Universit{\"a}t
Potsdam, Am Neuen Palais 10, D-14469 Potsdam, Germany. \par
 \noindent 
E-mail: hjschmi@rz.uni-potsdam.de \quad
  http://www.physik.fu-berlin.de/\~{}hjschmi}}
\newpage

\section{Introduction}

Conformal properties for theories of gravity appear on several different 
 levels. The best known example is the conformally coupled scalar
 field, where the conformal property is due to the fact, that in 
four dimensions, the operator $\Box - R/6$ is conformally invariant. 

\bigskip

Two other examples are: First, the Bach equation $B_{\a \b} =0$, 
 (see e.g. \cite{y0} and 
\cite{y1} and the   references there for details,  \cite{y16}, \cite{y15}
 for further recent approaches to the Bach equation, and \cite{y17} 
 for an overview on fourth-order gravity in the time period 
 1918 - 1990)  whose
 conformal invariance follows from the conformal invariance of 
the Weyl tensor $C^\a_{\ \b \g \d}$, and, second,  the equation of 
geometrodynamics, proposed in 1984 in \cite{y2}, see also 
\cite{y3} and \cite{y4}. As the details of the latter are not so 
well-known,  we sum up  their  properties in the Appendix.   

\bigskip

It is the purpose of the present paper to compare these two 
latter  approaches.

\section{Notation}

We apply the following notation: Riemann tensor:
\begin{equation}
\label{eq1}
R^\lambda  {}_{\sigma \alpha \beta}  = \G ^{\lambda}
_{\beta  \sigma ,\alpha}- \dots 
\end{equation}
Ricci tensor:
\begin{equation}
\label{eq2}
R_{\alpha \beta}  = R^\sigma  {}_{\    \a  \sigma  \beta}
\, .
\end{equation}

This is the rule 
for rearrangement of order of covariant differentiation:
\begin{equation}
\label{eq3}
Y^\lambda  {}_{;\alpha ;\beta}  = Y^\lambda  {}_{;\beta ;\alpha}  -
Y^{\sigma
}R^{\lambda}  {}_{\sigma \alpha \beta}  ;\quad Y_{\lambda ;\alpha ;\beta}  =
Y_{\lambda ;\beta ;\alpha}  + Y_{\sigma}  R^{\sigma}  {}_{\lambda \alpha
\beta } \,  .
\end{equation}
The Weyl tensor is:
\begin{equation}
\label{eq4}
\begin{array}{l}
C_{\alpha \beta \mu \nu}  \equiv
R_{\alpha \beta \mu \nu}  -
\frac{{1}}{{2}}\left[ {g_{\alpha \mu}  R_{\beta \nu}  + g_{\beta \nu}
R_{\alpha \mu}  - g_{\alpha \nu}  R_{\beta \mu}  - g_{\beta \mu}  R_{\alpha
\nu} }  \right] + \\
+\frac{{1}}{{6}}R\left[ {g_{\alpha \mu}  g_{\beta \nu}  -
g_{\alpha \nu}  g_{\beta \mu} }  \right] \, .
\end{array}
\end{equation}

\section{The Bach equation}

Bach equation describes the vanishing of the variational derivative of 
the square of the Weyl tensor with respect to the metric.
 We  use it here  as given in \cite{y1}, i.e.:
\begin{equation}
\label{eq5}
B_{\alpha \beta}  \equiv B_{\alpha \beta} ^{\left( {1} \right)} + B_{\alpha
\beta} ^{\left( {2} \right)} = 0  \,  .
\end{equation}
Here:
\begin{equation}
\label{eq6}
B_{\alpha \beta} ^{\left( {1} \right)} \left( {R_{ \cdot \cdot}  ;g_{ \cdot
\cdot} }  \right) = - R_{\alpha \beta} {}^{;\nu} {}_{;\nu}  +
R^{\nu}_{\a   ;\beta ;\nu} + R^{\nu}_{\b ;\alpha ;\nu}  -
\frac{{2}}{{3}}R_{;\alpha ;\beta}
+ \frac{{1}}{{6}}g_{\alpha \beta} R^{;\nu} {}_{;\nu} \,   ,
\end{equation}
\begin{equation}
\label{eq7}
B_{\alpha \beta} ^{\left( {2} \right)} \left( {R_{ \cdot \cdot}  ;g_{ \cdot
\cdot} }  \right) = \frac{{2}}{{3}}RR_{\alpha \beta}  - 2R_{\alpha \nu}
R^{\nu}{}_{\beta}  - \frac{{1}}{{6}}R^{2}g_{\alpha \beta}  +
\frac{{1}}{{2}}g_{\alpha \beta}  R_{\mu \nu}  R^{\mu \nu} \,  .
\end{equation}
From eqs. (\ref{eq6}), (\ref{eq7}) it becomes clear that 
 the quantities $B_{\alpha \beta} ^{\left( {1}
\right)}
,\;B_{\alpha \beta} ^{\left( {2} \right)} $  depend on the  Ricci tensor
and on  the metric. It should be mentioned, that up to a multiplication 
of the 
whole Bach tensor by a non-vanishing constant, all these 9 
coefficients in front of the 9 different geometric terms in 
 the right hand sides of eqs.  (\ref{eq6}) and (\ref{eq7})
 are uniquely determined by the conformal and covariant properties 
 of the Weyl tensor.

\section{The equation  of geometrodynamics}

By the ``equation  of geometrodynamics''   we denote  
the equation  deduced  in   \cite{y2}; it  reads (cf. the  Appendix
 for a deduction):
\noindent
\begin{equation}
\label{eq8}
R_{\alpha \beta}  - \frac{{1}}{{2}}g_{\alpha \beta}  R = - 2A_{\alpha}
A_{\beta}  - g_{\alpha \beta}  A^{2} - 2g_{\alpha \beta}  A^{\nu}
{}_{;\nu}  +
A_{\alpha ;\beta}  + A_{\beta ;\alpha} \, ,
\end{equation}
where
$$
A^2 = A_\m A_\n g^{\m \n} \, .
$$
From the equation (\ref{eq8}) we get after taking the divergence 
\begin{equation}
\label{eq9}
F_{\alpha} {}^{\nu} {}_{;\nu}  = 0 \,  ,
\end{equation}
\noindent
where
\noindent
\begin{equation}
\label{eq10}
F_{\alpha \beta}  \equiv A_{\beta ,\alpha}  - A_{\alpha ,\beta}  
 =  A_{\beta ;\alpha}  - A_{\alpha ; \beta}   \,  .
\end{equation}
\noindent
 The trace  of  equation (\ref{eq8})  yields:
\begin{equation}
\label{eq12}
R = 6A^{2} + 6A^{\nu}  {}_{;\nu} \,   .
\end{equation}
Then the  equation (\ref{eq8}) can also be written as
\begin{equation}
\label{eq11}
R_{\alpha \beta}  = - 2A_{\alpha}  A_{\beta}  + 2g_{\alpha \beta}  A^{2} +
g_{\alpha \beta}  A^{\nu}  {}_{;\nu}  + A_{\alpha ;\beta}  + A_{\beta
;\alpha}   \,   .
\end{equation}
\noindent

\section{The  definition of the problem}

We  apply the following conformal transformation:
\begin{equation}
\label{eq13}
g_{\alpha \beta}  \left( {x} \right) \to {g'}_{\alpha \beta}  \left( {x}
\right) = g_{\alpha \beta}  \left( {x} \right) \cdot e^{2\sigma \left( {x}
\right)} \, .
\end{equation}

\noindent
The equation of geometrodynamics (\ref{eq8}) is invariant with respect to
conformal
transformations (\ref{eq13}), if the vector $A_{\alpha}  \left( {x} \right)$
is
simultaneously transformed as:
\begin{equation}
\label{eq14}
A_{\alpha}  \left( {x} \right) \to {A'}_{\alpha}  \left( {x} \right) =
A_{\alpha}  \left( {x} \right) - \sigma _{,\alpha}  \left( {x} \right) \, .
\end{equation}

\bigskip

The Bach equation (\ref{eq5}) has the following solutions:

\medskip

$ - $ Riemannian spaces with zero Ricci tensor, i.e. with $R_{\alpha \beta}
= 0$. After a conformal transformation we get 
\begin{equation}
\label{eq15}
R_{\alpha \beta}  = - 2\varphi _{,\alpha}  \varphi _{,\beta}  + 2g_{\alpha
\beta}  g^{\mu \nu} \varphi _{,\mu}  \varphi _{,\nu}  + g_{\alpha \beta}
\varphi^{;\nu} {}_{;\nu} + \varphi _{;\alpha ;\beta}  + \varphi _{;\beta
;\alpha} \,  .
\end{equation}

\noindent
Here $\varphi \left( {x} \right)$ is 
an arbitrary function of the coordinates.
Due  to the conformal invariance of the Bach equation, 
Riemannian spaces with Ricci tensor
(\ref{eq15}) also satisfy the Bach equation. \footnote{Hint for a proof: 
Insert $\varphi = \sigma $  into eq.  (\ref{eq13}), then 
 eq.  (\ref{eq15}) implies $R'_{\a\b}\equiv 0$. The last two terms of  
eq.  (\ref{eq15}) coincide, 
of course, but we wrote them this way we did to ease the
 comparison with eq.  (\ref{eq11}).} 

\medskip

$ - $ Einstein spaces,

\begin{equation}
\label{eq16}
R_{\alpha \beta}  = \lambda \cdot g_{\alpha \beta}  ,
\quad
\lambda \ne 0,
\quad
\lambda = \  {\rm const.}   \   ,
\end{equation}

\noindent
as well as Riemannian spaces, that differ from the Einstein spaces by the
conformal transformation (\ref{eq13}). If this $\sigma$
is a constant function $\sigma =
\sigma  _{0} =$ const., then we get again an Einstein space, this  
time with ${\lambda}' = \lambda e^{ - 2\sigma _{0}} $.\footnote{Solutions
  of the Bach equation which are not conformally related to an Einstein 
space are 
called ``non-trivial solutions'', there existence is shown in \cite{y0},
 \cite{y16} and \cite{y15}.}  

\bigskip

As a result, we have two types of conformally invariant equations that
describe Riemannian space dynamics. Each type of equations can be treated as
a basis for a conformally invariant theory of gravitation. These theories
are  not equivalent, since there are the following differences between Bach
equations and equations of geometrodynamics:

\medskip
\noindent 
1. The equation (\ref{eq5}) is of the 4$^{th}$ order, while (\ref{eq8}) is a
second-order  equation.

\medskip
\noindent 
2. The equation (\ref{eq8}) includes the vector $A_{\alpha}  \left( {x}
\right)$,  and the eq. (\ref{eq5}) does not.

\medskip
\noindent 
3. For the equation (\ref{eq8}) a correctly-defined Cauchy problem exists
(see \cite{y3}).
As for the equation (\ref{eq5}), correctness of the Cauchy problem
definition  has  not been proved yet.\footnote{More concretely, there 
 are at least two open questions in this respect: First, that point in the 
usual Cauchy problem formulation of  the Bach equation, where one
 breaks the conformal invariance by requiring the curvature scalar 
to be a constant: it is not clear whether this restricts globally the set of 
solutions  or not; second, even for the simplest anisotropic cosmological 
models, namely Bianchi type I, it is not clear whether the statement 
``Without loss of generality, one may assume the Bianchi type I metric 
in diagonal form to get the full set of Bianchi type I vacuum solutions.'' 
proven for Einstein's theory, is valid for the Bach equation, too. This
 second question shall be discussed more detailed in a further paper.}

\medskip
\noindent 
4. The equation (\ref{eq5}) can be derived from the variation principle; no
Lagrangian with the dimension of $\left[ {length^{2}} \right]$ exists for
the equation (\ref{eq8}).

Remember that the Lagrangian for the Bach equation reads:

\begin{equation}
\label{eq17}
L = C_{\alpha \beta \mu \nu}  C^{\alpha \beta \mu \nu}  \,  .
\end{equation}

Taking into account the fact that
\begin{equation}
\label{eq18}
L_{GB} = R_{\alpha \beta \mu \nu}  R^{\alpha \beta \mu \nu}  - 4R_{\alpha
\beta}  R^{\alpha \beta}  + R^{2}
\end{equation}

\noindent
is the complete Gauss-Bonnet divergence, the Lagrangian for the equation
(\ref{eq5})
can also be written as \footnote{up to surface terms in the action which 
do not 
 influence the classical field equation}:
\begin{equation}
\label{eq19}
\tilde {L} = 2R_{\alpha \beta}  R^{\alpha \beta}  - \frac{{2}}{{3}}R^{2} \,  .
\end{equation}
It is clear that any solution of   the 
 Bach equation in the form (\ref{eq15}) with
a  constant $\lambda = 0$ \footnote{ The equation (\ref{eq8}) can  also be 
written with
a non-zero $\lambda $-term, and with 
 the transformations
(\ref{eq13}), (\ref{eq14}),  $\lambda$  will
be transformed as $\lambda \to {\lambda}' = \lambda \cdot e^{ - 2\sigma} $.
For our proof it is, however, sufficient to assume $\lambda = 0$.} will also
satisfy the equation (\ref{eq8}), since (\ref{eq15}) can be reduced to
(\ref{eq8}) by equating
\begin{equation}
\label{eq20}
A_{\alpha}  \left( {x} \right) = - \varphi _{,\alpha}  \left( {x} \right)
\quad .
\end{equation}
The subject of this paper is the 
 solution of the inverse problem, i.e. to find
out under what conditions solutions
 of the equation of geometrodynamics
(\ref{eq8})
are also solutions to the  Bach equation (\ref{eq5}).

\section{Solution of the problem}

The Ricci tensor (\ref{eq11}) will 
be referred to as geometrodynamical tensor   and
denoted as $R_{ \cdot \cdot} ^{\rm  GD} $. We will 
show that for $R_{ \cdot  \cdot} ^{\rm   GD} $ there exists
 the following relation:
\begin{equation}
\label{eq21}
B_{\alpha \beta}  \left( {R_{ \cdot \cdot} ^{\rm GD} ;g_{ \cdot \cdot} }
\right) = 2F_{\alpha \nu}  F_{\beta}  {}^\nu - \frac{{1}}{{2}}g_{\alpha
\beta}  F_{\mu \nu}  F^{\mu \nu} .
\end{equation}
The proof is lengthy. To make the procedure convenient for check, we will
divide the proof into 7 steps.

Step 1.

In this step $R_{ \cdot \cdot} ^{\rm GD} $ from (\ref{eq11}) is 
substituted to
$B_{\alpha
\beta} ^{\left( {1} \right)} \left( {R_{ \cdot \cdot}  ;g_{ \cdot \cdot} }
\right)$. This yields:
\begin{equation}
\label{eq22}
B_{\alpha \beta} ^{\left( {1} \right)} \left( {R_{ \cdot \cdot} ^{\rm GD}
{ 
\cdot \cdot} }  \right) = \Phi _{\alpha \beta}  \left( {{A}'''} \right) + 
\Phi _{\alpha \beta}  \left( {A{A}''} \right) + \Phi _{\alpha \beta}  \left( 
{{A}'{A}'} \right).
\end{equation}

The terms' dependence on the vector $A_{\alpha}  $ and its derivatives 
$A_{\alpha ;\beta}  $ is shown symbolically in brackets. In the explicit 
form, these terms read:
\begin{equation}
\label{eq23}
\begin{array}{l}
 \Phi _{\alpha \beta}  \left( {{A}'''} \right) = - A^{\nu}  {}_{;\nu ;\alpha 
;\beta}  - A^{\nu}  {}_{;\nu ;\beta ;\alpha}  + \\ 
 + A_{\alpha} {}^{;\nu} {}_{;\beta ;\nu}  + A_{\beta}   {}^{;\nu}
{}_{;\alpha 
;\nu}  
+ A^{\nu} {}_{;\alpha ;\beta ;\nu}  + A^{\nu} {}_{;\beta ;\alpha ;\nu}  - 
A_{\alpha ;\beta} {}^{;\nu} {}_{;\nu}  - A_{\beta ;\alpha} {}^{;\nu}
{}_{;\nu}    
\\ 
 \end{array}.
\end{equation}
\begin{equation}
\label{eq24}
\begin{array}{l}
 \Phi _{\alpha \beta}  \left( {A{A}''} \right) = 2\left( {A_{\alpha}
{}^{;\nu} 
{}_{;\nu}  }  
\right)A_{\beta}  + 2\left( {A_{\beta} {}^{;\nu} {}_{;\nu}  }
\right)A_{\alpha}  
- 2g_{\alpha 
\beta}  A^{\sigma} \left( {A_{\sigma} {}^{;\nu} {}_{;\nu}  }  \right) - \\ 
 - 2A_{\alpha}  A^{\nu} _{;\beta ;\nu}  - 2A_{\beta}  A^{\nu} _{;\alpha ;\nu 
} - 2A^{\nu} A_{\alpha ;\beta ;\nu}  - 2A^{\nu} A_{\beta ;\alpha ;\nu}  \\ 
 \end{array}
\end{equation}
\begin{equation}
\label{eq25}
\begin{array}{l}
 \Phi _{\alpha \beta}  \left( {{A}'{A}'} \right) = 4A_{\alpha ;\nu}  
A_{\beta}   {}^{;\nu}  - 2A_{\alpha ;\nu} 
 A^{\nu}  {}_{;\beta}  - 2A_{\beta
;\nu
} A^{\nu}   {}_{;\alpha}  - \\
 - 2A_{\alpha ;\beta}  \left( {A^{\nu}  {}_{;\nu} }  \right) - 2A_{\beta
;\alpha}  \left( {A^{\nu}  {}_{;\nu} }  \right) - 2g_{\alpha \beta}
\left( 
{A^{\mu ;\nu} A_{\mu ;\nu} }  \right) \\ 
 \end{array}
\end{equation}

Step 2.

In this step, the term $\Phi _{\alpha \beta}  \left( {{A}'''} \right)$ is 
first transformed with the Bianchi identity  to the following form:
\begin{equation}
\label{eq26}
\Phi _{\alpha \beta}  \left( {{A}'''} \right) = 2A^{\sigma} R_{\alpha \beta 
;\sigma} ^{\rm GD} + 2A^{\nu}  {}_{;\alpha}  R_{\nu \beta} ^{\rm GD} +
2A^{\nu 
}  {}_{;\beta}  R_{\nu \alpha} ^{\rm GD} ,
\end{equation}

\noindent
then, after the expression for $R_{ \cdot \cdot} ^{\rm GD} $ has been 
substituted, it reads:
\begin{equation}
\label{eq27}
\Phi _{\alpha \beta}  \left( {{A}'''} \right) = \Theta _{\alpha \beta}  
\left( {A{A}''} \right) + \Theta _{\alpha \beta}  \left( {{A}'{A}'} \right) 
+ \Theta _{\alpha \beta}  \left( {A^{2}{A}'} \right),
\end{equation}

\noindent
where
\begin{equation}
\label{eq28}
\Theta _{\alpha \beta}  \left( {A{A}''} \right) = 2g_{\alpha \beta}  A^{\nu 
}A^{\sigma}  {}_{;\sigma ;\nu}  + 2A^{\nu} A_{\alpha ;\beta ;\nu}  + 2A^{\nu 
}A_{\beta ;\alpha ;\nu}  ,
\end{equation}
\begin{equation}
\label{eq29}
\begin{array}{l}
 \Theta _{\alpha \beta}  \left( {{A}'{A}'} \right) = 2A_{\alpha ;\beta}  
\left( {A^{\nu}  {}_{;\nu} }  \right) + 2A_{\beta ;\alpha}  \left( {A^{\nu 
}  {}_{;\nu} }  \right) + 4A_{\nu ;\alpha}  A^{\nu}  {}_{;\beta}  + \\ 
 + 2A_{\nu ;\alpha}  A_{\beta}   {}^{;\nu}  + 2A_{\nu ;\beta}  A_{\alpha}  
   {}^{;\nu}  \\ 
 \end{array} \quad ,
\end{equation}
\begin{equation}
\label{eq30}
\begin{array}{l}
 \Theta _{\alpha \beta}  \left( {A^{2}{A}'} \right) = - 4A^{\nu} A_{\alpha 
;\nu}  A_{\beta}  - 4A^{\nu} A_{\beta ;\nu}  A_{\alpha}  + 8g_{\alpha \beta 
} \left( {A^{\mu} A^{\nu} A_{\mu ;\nu} }  \right) - \\ 
 - 4A^{\nu} A_{\nu ;\alpha}  A_{\beta}  - 4A^{\nu} A_{\nu ;\beta}  + 
4A_{\alpha ;\beta}  A^{2} + 4A_{\beta ;\alpha}  A^{2} \\ 
 \end{array} \quad .
\end{equation}

Step 3.

In this step, all terms are substituted to the relation (\ref{eq22}) 
for
$$
B_{\alpha \beta} ^{\left( {1} \right)}
 \left( {R_{ \cdot \cdot} ^{\rm GD}
;g_{ \cdot \cdot} }  \right) \,  ;
$$
here,  $\Phi _{\alpha \beta}  \left( {{A}'''}
\right)$ being
written in the form (\ref{eq27}), where the relations for
$$
\Theta
_{\alpha \beta} \left( {A{A}''} \right)\, ; \quad
  \Theta _{\alpha \beta}
 \left(
{{A}'{A}'} \right) \, ;
$$
and 
$\Theta _{\alpha \beta}  \left( {A^{2}{A}'} \right)$
are given by (\ref{eq28}), (\ref{eq29}), (\ref{eq30}), respectively. This 
yields:
\begin{equation}
\label{eq31}
B_{\alpha \beta} ^{\left( {1} \right)} \left( {R_{ \cdot \cdot} ^{\rm GD}
 ;g_{ 
\cdot \cdot} }  \right) = \Psi _{\alpha \beta}  \left( {A{A}''} \right) + 
\Psi _{\alpha \beta}  \left( {A^{2}{A}'} \right) + \Psi _{\alpha \beta}  
\left( {{A}'{A}'} \right),
\end{equation}

\noindent
where
\begin{equation}
\label{eq32}
\begin{array}{l}
 \Psi _{\alpha \beta}
  \left( {A{A}''} \right) = 2\left( {A_{\alpha} {}^{;\nu} 
{}_{;\nu}  }  
\right)A_{\beta}
  + 2\left( {A_{\beta} {}^{;\nu} {}_{;\nu}  }  \right)A_{\alpha}  
- 2g_{\alpha 
\beta}  A^{\sigma} \left( {A_{\sigma} {}^{;\nu} {}_{;\nu}  }  \right) - \\ 
 - 2A_{\alpha} 
 A^{\nu}  {}_{;\beta ;\nu}  - 2A_{\beta}  A^{\nu}  {}_{;\alpha 
;\nu 
} + 2g_{\alpha \beta}  A^{\sigma} A^{\nu}  {}_{;\nu ;\sigma}  \\ 
 \end{array} \quad ,
\end{equation}
\begin{equation}
\label{eq33}
\begin{array}{l}
 \Psi _{\alpha \beta}  \left( {A^{2}{A}'} \right) = 4A_{\alpha ;\beta}  
A^{2} + 4A_{\beta ;\alpha}  A^{2} + 8g_{\alpha \beta}  \left( {A^{\mu 
}A^{\nu} A_{\mu ;\nu} }  \right) - \\ 
 - 4A^{\nu} A_{\alpha ;\nu} 
 A_{\beta}  - 4A^{\nu} A_{\beta ;\nu}  A_{\alpha 
} - 4A^{\nu} A_{\nu ;\alpha}  A_{\beta}  - 4A^{\nu} A_{\nu ;\beta}  
A_{\alpha}  \\ 
 \end{array} \quad ,
\end{equation}
\begin{equation}
\label{eq34}
\Psi _{\alpha \beta}  \left( {{A}'{A}'} \right) =
4A_{\alpha ;\nu}  A_{\beta
}   {}^{;\nu}  + 2A_{\nu ;\alpha}  A^{\nu}   {}_{;\beta}  - 2g_{\alpha
\beta}
\left( {A^{\mu ;\nu} A_{\mu ;\nu} }  \right) \quad .
\end{equation}

In this step, 
the terms with the third-order derivatives disappear from the
 relation for $B_{\alpha \beta} ^{\left( {1} \right)} \left( {R_{ \cdot
\cdot} ^{\rm GD} ;g_{ \cdot \cdot} }  \right)$. 
The next step will be to exclude
the second-order derivatives from $B_{\alpha \beta} ^{\left( {1} \right)}
\left( {R_{ \cdot \cdot} ^{\rm GD} ;g_{ \cdot \cdot} }  \right)$.

Step 4.

We transform the expression $\Psi _{\alpha \beta}  \left( {A{A}''} \right)$,
using the properties of the Riemann tensor. This yields:
\begin{equation}
\label{eq35}
\begin{array}{l}
 \Psi _{\alpha \beta}  \left( {A{A}''} \right) = - 2g_{\alpha \beta}
A^{2}\left( {A^{\nu}  {}_{;\nu} }  \right) - 4g_{\alpha \beta}
\left( {A^{\mu
}A^{\nu} A_{\mu ;\nu} }  \right) - \\
 - 2A_{\alpha}  F_{\beta}  {}^{\nu} {}_{;\nu}  - 2A_{\beta}  F_{\alpha}
{}^{\nu
} {}_{;\nu}  + 2g_{\alpha \beta}  A^{\sigma} F_{\sigma}  {}^{\nu} {}_{;\nu}
\\
 \end{array} \quad .
\end{equation}

If the Ricci tensor is written as (\ref{eq11}), then the vector 
$A_{\alpha}$ shall
necessarily satisfy the equation (\ref{eq9}). Taking this into account,
(\ref{eq35}) will
change to
\begin{equation}
\label{eq36}
\Psi _{\alpha \beta}  \left( {A{A}''} \right) = - 2g_{\alpha \beta}
A^{2}\left( {A^{\nu}  {}_{;\nu} }  \right) - 4g_{\alpha \beta}
\left( {A^{\mu
}A^{\nu} A_{\mu ;\nu} }  \right) \quad .
\end{equation}

Note that at this step the terms with second-order derivatives have been
replaced by the terms with first-order derivative.

Step 5.

We will find the resulting relation for $B_{\alpha \beta} ^{\left( {1}
\right)} \left( {R_{ \cdot \cdot} ^{\rm GD} ;g_{ \cdot \cdot} }  \right)$.
\begin{equation}
\label{eq37}
B_{\alpha \beta} ^{\left( {1} \right)} \left( {R_{ \cdot \cdot} ^{\rm GD}
{ 
\cdot \cdot} }  \right) = \Omega _{\alpha \beta}  \left( {A^{2}{A}'} \right) 
+ \Omega _{\alpha \beta}  \left( {{A}'{A}'} \right),
\end{equation}

\noindent
where
\begin{equation}
\label{eq38}
\begin{array}{l}
 \Omega _{\alpha \beta}  \left( {A^{2}{A}'} \right) = 4A_{\alpha ;\beta}  
A^{2} + 4A_{\beta ;\alpha}  A^{2} + 4g_{\alpha \beta}  \left( {A^{\mu 
}A^{\nu} A_{\mu ;\nu} }  \right) - \\ 
 - 4A^{\nu} A_{\alpha ;\nu}  A_{\beta}  - 4A^{\nu} A_{\beta ;\nu}  A_{\alpha 
} - 4A^{\nu} A_{\nu ;\alpha}  A_{\beta}  - 4A^{\nu} A_{\nu ;\beta}  
A_{\alpha} 
 - 2g_{\alpha \beta}  A^{2}\left( {A^{\nu}  {}_{;\nu} }  \right) \\ 
 \end{array}  ,
\end{equation}
\begin{equation}
\label{eq39}
\Omega _{\alpha \beta}  \left( {{A}'{A}'} \right) = 4A_{\alpha ;\nu}  
A_{\beta}   
{}^{;\nu}  + 4A_{\nu ;\alpha}  A^{\nu}  {}_{;\beta}  - 2g_{\alpha 
\beta}  \left( {A^{\mu ;\nu} A_{\mu ;\nu} }  \right) \quad .
\end{equation}

Step 6.

We substitute the 
expression for $R_{ \cdot \cdot} ^{\rm GD} $ in the relation 
for 
$B_{\alpha \beta} ^{\left( {2} \right)} 
\left( {R_{ \cdot \cdot} ^{\rm GD} ;g_{ \cdot \cdot} }  \right)$. 
All algebraic terms (the terms that do not 
include derivative of the vector $A_{\alpha}  $) are eliminated, 
only terms 
with the derivatives of not higher than the first order remaining. 

\begin{equation}
\label{eq40}
B_{\alpha \beta} ^{\left( 
{2} \right)} \left( {R_{ \cdot \cdot} ^{\rm GD} ;g_{ 
\cdot \cdot} }  \right) = \Xi _{\alpha \beta}  \left( {A^{2}{A}'}
 \right) + 
\Xi _{\alpha \beta}  \left( {{A}'{A}'} \right),
\end{equation}

\noindent
here
\begin{equation}
\label{eq41}
\Xi _{\alpha \beta}  \left( {A^{2}{A}'} \right) = -
 \Omega _{\alpha \beta}  
\left( {A^{2}{A}'} \right) \quad ,
\end{equation}
\begin{equation}
\label{eq42}
\begin{array}{l}
 \Xi _{\alpha \beta}  \left( {{A}'{A}'} \right) = - 2A_{\alpha ;\nu}  
A_{\beta}   {}^{;\nu}  - 2A_{\alpha ;\nu}  A^{\nu}   {}_{;\beta}  - 
2A_{\beta 
;\nu 
} A^{\nu}  {}_{;\alpha}  \\ 
 - 2A_{\nu ;\alpha}  A^{\nu}  {}_{;\beta}  + g_{\alpha \beta} 
 \left( {A^{\mu 
;\nu} A_{\mu ;\nu} }  \right) + g_{\alpha
\beta}  \left( {A^{\mu ;\nu
}A_{\nu ;\mu} }  \right) \\
 \end{array} \quad .
\end{equation}

Step 7.

The proof of the relation (\ref{eq21}). To prove this relation, 
we must
substitute the 
relation (\ref{eq37}) for $B_{\alpha \beta} ^{\left( {1}
\right)}
\left( {R_{ \cdot \cdot} ^{\rm GD} ;g_{ \cdot \cdot} }  \right)$ 
and the
relation (\ref{eq40}) for  $B_{\alpha \beta} ^{\left( {2} \right)}
\left( {R_{ 
\cdot \cdot} ^{\rm GD} ;g_{ \cdot \cdot} }  \right)$ 
into the RHS of (\ref{eq21}).
 So, the relation (\ref{eq21}) has been proven.
  The relation (\ref{eq21}) leads  to the following
 three statements.

\underline {Statement 1}:

If $A_{\alpha}  $ is 
a gradient vector, i.e. it is in the form (\ref{eq20}), 
then 
the Riemannian space with 
Ricci tensor (\ref{eq11}) is a solution to the equation 
(\ref{eq5}), since in this case $F_{\alpha \beta}  = 0$.

\underline {Statement 2}:

If we take a Lagrangian in the 
form\footnote{ This Lagrangian will be 
referred to as Weyl-Maxwell Lagrangian (WM Lagrangian), see
 also \cite{y5} for it.}  (43) instead 
of (\ref{eq19}),
\begin{equation}
\label{eq43}
\tilde {\tilde {L}} = \tilde {L} + 2F_{\mu \nu}  F^{\mu \nu}  = 
2R_{\alpha 
\beta}  R^{\alpha \beta}  - \frac{{2}}{{3}}R^{2} + 2F^{2},
\end{equation}

\noindent
then the generalized Bach equation will read
\begin{equation}
\label{eq44}
{B}'_{\alpha \beta}  \equiv B_{\alpha \beta} ^{\left( {1} \right)} + 
{B'}_{\alpha \beta}^{\left( {2} \right)} = 0,
\end{equation}

\noindent
where $B_{\alpha \beta} ^{\left( {1} \right)} $ is given by (\ref{eq6}), 
and 
${B'}_{\alpha \beta}^{\left( {2} \right)} $ is as follows:
\begin{equation}
\label{eq45}
\begin{array}{l}
 {B'}_{\alpha \beta}^{\left( {2} \right)} 
= B_{\alpha \beta} ^{\left( {2} 
\right)} - 2F_{\alpha \nu}  F^{\nu} {}_{\beta} 
 - \frac{{1}}{{2}}g_{\alpha 
\beta}  \left( {F_{\mu \nu}  F^{\mu \nu} } \right)  
 = \frac{{2}}{{3}}RR_{\alpha \beta}  -\\
- 2R_{\alpha \nu}  R^{\nu} {}_{\beta}  - 
\frac{{1}}{{6}}R^{2}g_{\alpha \beta}  + \frac{{1}}{{2}}g_{\alpha \beta}  
R_{\mu \nu}  R^{\mu \nu}  - 2F_{\alpha \nu}  F^{\nu} {}_{\beta}  - 
\frac{{1}}{{2}}g_{\alpha \beta}  
\left( {F_{\mu \nu}  F^{\mu \nu} } \right) 
\\ 
 \end{array}
\end{equation}
The Ricci tensor (\ref{eq11}) satisfies the equation (\ref{eq44}). 

\underline {Statement 3}:

If the Lagrangian (\ref{eq19}) is replaced by the Lagrangian
\begin{equation} \label{eq46}
\tilde {L} + {\rm 
const}  \cdot F_{\mu \nu}  F^{\mu \nu}  = 2R_{\alpha \beta}  
R^{\alpha \beta}  - \frac{{2}}{{3}}R^{2} + {\rm const}  \cdot F^{2},
\end{equation}

\noindent
where const $  \ne + 2$, then 
the Ricci tensor (\ref{eq11}) will only satisfy the 
dynamic equation if the vector $A_{\alpha}  $ is in the gradient form.

\section{The spherically symmetric solutions  of the  Bach equation 
and 
 the geometrodynamics equation}

Any spherically symmetric solution of the 
Bach equation is equivalent (almost everywhere and up to   a 
conformal factor) to the Schwarzschild  static  solution in  
general 
relativity (GR). Each of these solutions in  GR is governed by a single 
parameter, i.e. the gravitational radius. 

For the geometrodynamics equation (\ref{eq8}) a spherically symmetric 
solution 
can 
also be given in the  static  form. However, the spherically symmetric 
solution is now governed by two parameters, rather than one. We will 
write 
explicitly one of the possible static forms of the spherically symmetric 
solutions \footnote{ A detailed description of how to derive different 
forms 
of spherically symmetric solutions to geometrodynamic equation will be 
covered in a separate publication.} of the equation (\ref{eq8}). 
The peculiarity 
of 
this form is that the radial component is also the one that describes 
brightness.
\begin{equation}
\label{eq47}
\begin{array}{l}
 ds^{2} =
 - \left(  {1 + \frac{{r}}{{R_{0}} }}  \right)  
   \left( {1 -   \frac{{r}}{{R_{0}} }}   \right)
  \left( {1 - \frac{{r_{0}} }{{r}}} \right)
  \cdot 
dt^{2} +
 \frac{dr^2}
{{\left( {1 + \frac{{r}}{{R_{0}} }} \right)\left( {1 
- \frac{{r}}{{R_{0}} }} \right)
\left( {1 - \frac{{r_{0}} }{{r}}} \right)}} +
\\
 + r^{2} \cdot \left[ {d\theta ^{2} + \sin^{2}\theta \cdot d\varphi ^{2}}
\right] \,   .   \\
 \end{array} 
\end{equation}

At $r_{0} < r < < R_{0} $ the metric (\ref{eq47}) is similar to the 
Schwarzschild
metric,
where $r_{0} $ works as a gravitational radius. It holds: For $R_0=0$, 
metric (\ref{eq47}) is the exact Schwarzschild solution, for $r_0=0$, 
metric (\ref{eq47}) is the 
exact de Sitter  solution, but for $r_0 \cdot R_0 \ne 0$,
 it does {\it not} give the Schwarzschild-de Sitter solution.

Meanwhile, the topology
 of the solution
  (\ref{eq47}) is qualitatively different from the topology of the 
Schwarzschild 
solution.  The difference is,  that the solution (\ref{eq47}) has a
 horizon  not  only at the surface
\begin{equation}
\label{eq48}
r = r_{0} \,  ,
\end{equation}

\noindent
but also a singularity of the conformal  curvature 
invariant 
at the surface
\begin{equation}
\label{eq49}
r = R_{0} \,  .
\end{equation}

\section{Discussion}

We must point out some facts given by the obtained results.

$ \bullet $ This paper illustrated, with a  static  spherically symmetric 
solution as an example, the correspondence between the solution of the  
Bach 
equation 
(the solution with the vector $A_{\alpha}  $ given by $A_{\alpha}  
= - \varphi _{,\alpha}  $) and that of   the geometrodynamics equation 
(i.e. the 
solution with non-gradient vector $A_{\alpha}  $). The fact is that a 
solution involving 
a non-gradient vector $A_{\alpha}  $ ($A_{\alpha}  \ne - 
\varphi _{,\alpha}  $) cannot coincide, in principle, with the 
Schwarzschild 
solution in the entire range of variation of variables, though in some 
interval of   the  radial coordinate it may come close to this solution. 

$ \bullet $ Solutions of eq.  (\ref{eq8}) describe  dissipative 
phenomena like 
heat 
conduction and viscosity (see \cite{y4}). One of the results of 
this paper is the 
proven fact that solutions of this type are also solutions to 
the equation 
(\ref{eq44}), that can be derived from the holonomic 
variation principle by a 
conventional procedure. This fact itself is non-trivial. Another 
amazing 
feature is that the structure of the corresponding holonomic 
Lagrangian is 
unambiguous. 

$ \bullet $ It is likely that the structure of the  Weyl-Maxwell 
 Lagrangian in the form 
(\ref{eq43}) enables the 
construction of an analogue of a  Hamiltonian formalism for solutions 
of the 
equation (\ref{eq8}) and to quantize, despite the fact that, 
this equation 
describes 
dissipative phenomena. 

$ \bullet $ The addition of the Maxwellian 
term $\left( {F_{\mu \nu}  F^{\mu \nu 
}} \right)$ to the Lagrangian (\ref{eq43}) does not 
 immediately  allow to interpret the  vector 
$A_{\alpha}  $ as an  electromagnetic vector potential. The Lagrangian 
(\ref{eq43}) 
does not yield the  Einstein field  equations with the Maxwell
 energy-momentum
tensor, but the equation (\ref{eq44}). The Lagrangian (\ref{eq43}) 
includes
two  conformally
invariant scalar values with Weyl weights, inversely proportional 
to $\sqrt
{ - g} $. These scalars are: $ {C_{\alpha \beta \mu \nu}  C^{\alpha
\beta \mu \nu} } $ and $ {F_{\mu \nu}  F^{\mu \nu} } $.
They give a  complete list of the Weyl space invariants \footnote{ The
quantities $ {E_{\alpha \beta \mu \nu}  C^{\alpha \beta \sigma \rho
}C_{\sigma \rho}  {}^{\mu \nu} } $ and  $  {E_{\alpha \beta \mu \nu}
F^{\alpha \beta} F^{\mu \nu} }$ have the same Weyl  weight, they
are, however, pseudoscalars. Here, $E_{\cdot \cdot \cdot \cdot}$ denotes
the dual of the Weyl tensor.}. In fact, the Lagrangian that
includes both invariants is the only possible one among those able to
produce a dimensionless action with zero Weyl weight, in the case of
Lagrangian being constructed immediately for the Weyl space. For this 
reason
 this represents a more natural 
interpretation for the vector $A_{\alpha}  $ as being a Weyl
vector. 

\section*{Appendix: Deduction of the properties of geometrodynamics}

Here we give a deduction of  basic facts  connected with the
equation of geometrodynamics. This deduction is both a new  approach  
and
 also  more general than that ones given in the literature up to now.

Assume that the source for the Einstein field equation shall be 
composed 
 from a 
 vector field $A_\a$, where at most first derivatives of it shall 
enter the 
energy-momentum tensor. A possible $\Lambda$-term can  be omitted
 without loss of generality, \footnote{With $\Lambda \ne 0$ the 
result would be  analogous.}
and terms of cubic and higher degree in   $A_\a$ shall be omitted. 
Then we
are left with the following equation 
\be\label{t0}
 R_{ \a\b} - \frac{R}{2} g_{\a\b} = c_1 \, A_\a A_\b + c_2 \, 
A_{(\a ; \b)} + c_3 \,  g_{\a \b} A^2 + c_4 \, g_{\a \b} A^\g_{\  ;\g}
\ee
containing the four free parameters $c_i$, $i=1,  \dots , 4$. Here,
$A^2=A^\g A_\g$ and the round symmetrization bracket  is defined via 
\be\label{t1}
A_{(\a ; \b)} = \frac{1}{2} \left( A_{\a ; \b} + A_{\b ; \a}
\right) \, .
\ee 
This equation  (\ref{t0}) goes 
 over to eq. (\ref{eq8}), if the quadruple $c_i$ takes the form
\be\label{t2}
c_i = (-2, \, 2, \, -1, \, -2 \, ) \ .
\ee
The trace of   eq. (\ref{t0}) reads 
\be\label{t3}
-R=(c_1+4c_3)A^2 + (c_2 +4c_4) A^\g_{ \  ;\g} \,  .
\ee
If one inserts eq. (\ref{t2}) into eq. (\ref{t3}), one gets, of course, 
  eq. (\ref{eq12}). 

We take the covariant divergence of  eq. (\ref{t0}). By use of the Bianchi
identity, the l.h.s. identically vanishes, so we get
\be\label{t4}
0=c_1  A^\g_{\, ;\g}A_\b + c_1 A^\g A_{\b;\g} + \frac{c_2}{2} 
 A^\g _{\, ; \b \g} + \frac{c_2}{2} \Box A_\b + 2c_3 A^\g A_{\g;\b}
+ c_4 A^\g_{\, ; \g \b} \, .
\ee
For the further calculation it proves useful to define a second vector  field,
 $T_\b$, 
 via the equation 
\be\label{t5}
 T_\b = A^\a \, R_{\a \b} \, .
\ee
With  eqs. (\ref{t0}) and  (\ref{t3}) we find
\be\label{t6}
T_\b = c_2 A^\a A_{(\a;\b)} + A_\b
\left(
(\frac{c_1}{2} - c_3) A^2 - (\frac{c_2}{2} + c_4) A^\g_{\ ;\g}
 \right)   \, .
\ee
To keep simplicity, we again require the absence of cubic terms, i.e.
\be\label{t7}
c_1 = 2c_3 \, .
\ee
For the quadruple eq. (\ref{t2}) this condition is fulfilled.

Now we apply  eq. (\ref{eq3}) and find with eq. (\ref{t5})
\be\label{t8}
A^\a_{\  ;\b\a} - A^\a_{ \  ;\a\b} =  T_\b \, . 
\ee
Eq. (\ref{t6}) with eq. (\ref{t7}) gives 
\be\label{t9}
T_\b = c_2 A^\a A_{(\a;\b)} 
 - (\frac{c_2}{2} + c_4) A_\b  A^\g_{\ ;\g}   \, .
\ee
Eq. (\ref{t4}) with eq. (\ref{t7}) gives 
\be\label{t10}
0=2 c_3  A^\g_{\, ;\g}A_\b + 4c_3 A^\g A_{(\b;\g)}
 + \frac{c_2}{2} 
 A^\g _{\, ; \b \g} + \frac{c_2}{2} \Box A_\b  
+ c_4 A^\g_{\, ; \g \b} \, .
\ee
The goal is now to find out under which circumstances eq. (\ref{eq9})
 can be deduced from eqs. (\ref{t8}) till (\ref{t10}).
 Eq. (\ref{eq9}) with eq. (\ref{eq10}) reads 
\be\label{t11}
 \Box A_\b   = A^\g _{\, ; \b \g} \, .
\ee

Therefore, the term $A^\g_{\, ; \g \b}$ of eq.  (\ref{t10})
 is superfluous. We take the 
three   eqs. (\ref{t8}) till (\ref{t10}), cancel the vector $T_\b$ and the
 term $A^\g_{\, ; \g \b}$ and get afterwards:
\bea\label{t12}
0=2 c_3  A^\g_{\, ;\g}A_\b + 4c_3 A^\g A_{(\b;\g)}
 + \frac{c_2}{2} 
 A^\g _{\  ; \b \g} + \frac{c_2}{2} \Box A_\b + \nonumber \\
+ c_4 A^\g_{\, ; \b \g}
- c_2 c_4 A^\g A_{(\b;\g)} + c_4 \left( \frac{c_2}{2} + c_4 \right)
 A_\b A^\g_{\ ; \g} \, .
\eea
 To get a relation like    eq. (\ref{t11}), the nonlinear terms in 
    eq. (\ref{t12}) have to be cancelled. This takes place if 
\be\label{t13}
4 c_3= c_2 \cdot c_4
\ee
and 
\be\label{t14}
2 c_3 = -  c_4 \left( \frac{c_2}{2} + c_4 \right)   \, .
\ee
$c_2 =0$ implies $c_i=0$ for  all $i$, which is a non-interesting case.
 Therefore, we assume $c_2 \ne 0$.  We are free to redefine the vector 
 $A_\b$ by multiplying it with any non-vanishing constant, because  
 we can compensate this by an
 appropriate redefinition of the constants $c_i$, see eq. (\ref{t0}). We use 
this freedom to choose the value $c_2$ from eq. (\ref{t2}), i.e., 
$c_2=2$. Then eqs. (\ref{t13}),   (\ref{t14}) read:
$$
4c_3 = 2c_4 \, ; \qquad 2c_3 = - c_4(1+c_4) \, .
$$
The solution with $c_4=0$, which  implies also $c_1= c_3 =0$, turns 
eq. (\ref{t10}) into 
$$
 \Box A_\b   = - A^\g _{\, ; \b \g} \, .
$$
This special case shall now be omitted, then we get as the only other
 solution  $c_4 = -2$, $c_3= -1$, and with eq. (\ref{t7}) finally
 $c_1 = -2$. This is together just the original quadruple 
 eq. (\ref{t2}). So, we have shown, that eq. (\ref{t11}) indeed follows
 without further assumptions from the divergence of eq. (\ref{eq8}), and
 we have shown in which sense the coefficients $c_i$ are uniquely determined.

\section*{Acknowledgement}

This paper is part of the ISTC project KR 154 (Cosmological consequences of
 the 
quantum `birth' of the world).

\end{document}